\documentclass[aps,prc,preprint,amsmath,amssymb,showpacs,preprintnumbers,superscriptaddress]{revtex4-1}

\usepackage{CJK}
\usepackage{graphicx}
\usepackage{dcolumn}
\usepackage{bm}

\usepackage{color}
\usepackage{hyperref}

\usepackage{longtable}
\usepackage{mathrsfs}
\usepackage{multirow}

\begin{document}
\begin{CJK*}{UTF8}{}
%

\title{Covariant Density Functional Theory with Localized Exchange Terms}

\author{Qiang Zhao \CJKfamily{gbsn} (赵强)}
\affiliation{State Key Laboratory of Nuclear Physics and Technology, School of Physics, Peking University, Beijing 100871, China}
\affiliation{Center for Exotic Nuclear Studies, Institute for Basie Scicence, Daejeon 34126, Korea}

\author{Zhengxue Ren \CJKfamily{gbsn} (任政学)}
\affiliation{State Key Laboratory of Nuclear Physics and Technology, School of Physics, Peking University, Beijing 100871, China}

\author{Pengwei Zhao  \CJKfamily{gbsn} (赵鹏巍) }
\email{pwzhao@pku.edu.cn}
\affiliation{State Key Laboratory of Nuclear Physics and Technology, School of Physics, Peking University, Beijing 100871, China}

\author{Jie Meng  \CJKfamily{gbsn} (孟杰) }
\email{mengj@pku.edu.cn}
\affiliation{State Key Laboratory of Nuclear Physics and Technology, School of Physics, Peking University, Beijing 100871, China}
\affiliation{School of Physics and Nuclear Energy Engineering, Beihang University, Beijing 100191, China}
\affiliation{Yukawa Institute for Theoretical Physics, Kyoto University, Kyoto 606-8502, Japan}

\date{\today}

\begin{abstract}
A new density-dependent point-coupling covariant density functional PCF-PK1 is proposed, where the exchange terms of the four-fermion terms are local and are taken into account with the Fierz transformation.
The coupling constants of the PCF-PK1 functional are determined by empirical saturation properties and \emph{ab initio} equation of state and proton-neutron Dirac mass splittings for nuclear matter as well as the ground-state properties of selected spherical nuclei.
The success of the PCF-PK1 is illustrated with properties of the infinite nuclear matter and finite nuclei including the ground-state properties and the Gamow-Teller resonances.
In particular, the PCF-PK1 eliminates the spurious shell closures at $Z=58$ and $Z=92$, which exist commonly in many covariant density functionals without exchange terms.
Moreover, the Gamow-Teller resonances are nicely reproduced  without any adjustable parameters, and this demonstrates that a self-consistent description for the Gamow-Teller resonances can be achieved with the localized exchange terms in the PCF-PK1.
\end{abstract}

\maketitle
\end{CJK*}

\section{Introduction}

The development of rare isotope beam facilities has largely extended the limits of our exploration on the nuclear chart from the stability valley to the drip lines.
In particular, the unstable nuclei with large isospin could exhibit many unexpected phenomena, which considerably deepen our understanding of nuclear physics \cite{casten2000_PiPaNP45_S171, tanihata2013_PiPaNP68_215}.
In addition, the properties of unstable nuclei also provide essential inputs for the study of nucleosynthesis in the universe.
However, most of the unstable nuclei are still far beyond the current experimental capacities.
Therefore, it is important to build a unified and consistent nuclear many-body method to study the unstable nuclei and provide reliable descriptions.

The covariant density functional theory (DFT) is one of the most successful microscopic theories for a global description of nuclei~
\cite{Meng2016_IRNP10,meng2021_AB31_2}.
It is based on the relativistic quantum field theory and density functional theory for quantum many-body problems~\cite{Serot1986Adv.Nucl.Phys.1,Ring1996Prog.Part.Nucl.Phys.193,Vretenar2005Phys.Rep.101,Meng2006Prog.Part.Nucl.Phys.470}.
The covariant DFTs take Lorentz symmetry into account and provide an efficient description of nuclei with the underlying large scalar and vector fields of the order of a few hundred MeV, which are hidden in the nonrelativistic DFTs.
This is clearly seen in the nonrelativistic reduction of covariant DFT via a similarity renormalization method, which provides a possible means to bridge the relativistic and non-relativistic DFTs~\cite{Ren2020Phys.Rev.C21301}.
The inclusion of Lorentz symmetry brings several advantages for the covariant DFTs~\cite{Ring2012PhysicaScripta14035}.
For instance, it naturally includes the spin degree of freedom, and automatically gives the large spin-orbit potential in nuclei.
It provides a consistent treatment of the time-odd fields, which are particularly important for describing spectroscopic properties associated with nuclear rotations~\cite{Afanasjev2010Phys.Rev.C34329,Meng2013Front.Phys.55}.
Due to these advantages, the covariant DFTs have attracted a lot of attentions during the past decades, and have been successfully applied to study quantitatively the static and dynamic properties of nuclei~\cite{Ring1996Prog.Part.Nucl.Phys.193,Vretenar2005Phys.Rep.101,Meng2006Prog.Part.Nucl.Phys.470,Niksic2011Prog.Part.Nucl.Phys.519,Ren2020Phys.Lett.B135194}.

In most of these applications, however, only the direct (Hartree) terms are considered, and the exchange (Fock) terms are usually neglected.
In principle, at the mean-field level, both the direct and exchange diagrams should be included.
It was assumed that the effects of the exchange terms could be absorbed in a phenomenological way in the adjustments of the coupling constants of energy density functionals.
However, except for simplicity, there is no robust physical reason for this assumption.
In fact, the contribution of the pion mesons to the mean fields can be taken into account only via the exchange terms, due to its negative-parity nature.

The effects of the exchange terms have been discussed within the framework of the relativistic Hartree-Fock (RHF) approach~\cite{Bouyssy1987Phys.Rev.C380,Long2006Phys.Lett.B150,Long2010Phys.Rev.C24308}.
By introducing the exchange terms, the exchange of $\pi$ mesons as well as the $\rho$-tensor couplings can be taken into account \cite{Long2007Phys.Rev.C34314}, and the tensor forces are naturally included~\cite{Jiang2015Phys.Rev.C34326,Wang2018Phys.Rev.C34313}.
It was found that the RHF functionals with the $\pi$-exchange and $\rho$-tensor couplings could improve the descriptions of the single-particle energies~\cite{Long2008Europhys.Lett.12001,Long2009Phys.Lett.B428,Wang2013Phys.Rev.C47301,Li2016Phys.Lett.B97,Liu2020Phys.Lett.B135524}.
Another effect of the exchange terms was revealed in the description of spin-isospin excitations with the relativistic random phase approximation, where no additional parameters are needed with the inclusion of exchange terms~\cite{Liang2008Phys.Rev.Lett.122502,Liang2012Phys.Rev.C64302}, while the Migdal term has to be adjusted with only the direct terms.
However, the bottleneck is that the nonlocal potentials involved in the meson-exchange RHF framework leads to a surge in the computational cost and, thus, the application of RHF is usually limited to spherical nuclei and light deformed ones \cite{Ebran2011Phys.Rev.C64323}.
Only recently, the axially deformed RHF model expanded on Dirac-Woods-Saxon basis is established \cite{Geng2020Phys.Rev.C64302}, which is able to be applied for heavy nuclei but still with a large demand of computational costs.

On the other hand, the relativistic point-coupling model provides another way to construct a relativistic energy density functional~\cite{Nikolaus1992Phys.Rev.C1757,Burvenich2002Phys.Rev.C44308,Zhao2010Phys.Rev.C54319}, where the meson exchange is replaced by the corresponding local (contact) interactions between the nucleons, and the finite-range effects are approximated by local derivative terms.
Such a replacement can be justified since it is known from the meson-exchange models that the exchanged $\sigma$, $\omega$, and $\rho$ mesons between nucleons are all heavy mesons.
In recent years, the relativistic point-coupling model has attracted more and more attention owing to the following advantages.
First, it is considerably simpler in numerical applications by avoiding the solution of the Klein-Gordon equations~\cite{Zhao2011Phys.Lett.B181} and also the complicated two-body matrix elements of finite-range in applications going beyond static mean-field theory~\cite{Niksic2011Prog.Part.Nucl.Phys.519}.
Second, it provides more opportunities to investigate its relationship to the nonrelativistic approaches~\cite{Sulaksono2003Ann.Phys.354,Ren2020Phys.Rev.C21301}.
Third, it is relatively easy to include the exchange terms with the point-coupling interactions~\cite{Sulaksono2003Ann.Phys.36,Liang2012Phys.Rev.C21302}.

In the framework of the relativistic point-coupling models, it is convenient to use the Fierz transformation to express the exchange terms as superpositions of the direct terms, such that the exchange terms can be treated in a similar way as the direct terms.
In Ref.~\cite{Liang2012Phys.Rev.C21302}, by the zero-range reduction and the Fierz transformation,  the validity of the localized exchange terms has been studied, and it retains the simplicity of point-coupling models and provides proper descriptions of the spin-isospin excitations and the Dirac masses.
Inspired by these findings, the present work is to propose a practical point-coupling covariant density functional with localized exchange terms obtained with the Fierz transformation.

The paper is organized as follows.
The theoretical framework of the covariant DFT with localized exchange terms is presented in Sec. \ref{sec:framework}.
In Sec. \ref{sec:numerical} and \ref{sec:strategy}, the numerical details as well as the strategy to determine the parameters in the density functional are given, respectively.
Then, the performance of the new density functional is tested in Sec. \ref{sec:result} by studying properties of infinite nuclear matter and finite nuclei including the ground-state properties and the Gamow-Teller resonance excitations.
Finally, the summary is given in Sec. \ref{sec:summary}.

\section{Theoretical framework}\label{sec:framework}

The elementary building blocks of the relativistic point-coupling model are
\begin{align}
  (\bar\psi\mathcal{O}\Gamma\psi),\qquad
  \mathcal{O}\in\{1,\vec{\tau}\},\qquad
  \Gamma\in\{1,\gamma_\mu,\gamma_5,\gamma_5\gamma_\mu,\sigma_{\mu\nu}\},
\end{align}
where $\psi$ is the nucleon field, $\vec{\tau}$ is the isospin matrices, and $\Gamma$ represents the Dirac matrices.
Here, the arrows are used to denote vectors in the isospin space, and the space vectors will be denoted as the bold types.

The Lagrangian density of the nuclear system can be constructed as the power series of these building blocks and their derivatives.
The starting point is an effective Lagrangian density consisting of four parts,
\begin{align}\label{equ:Lagrangian}
  \mathcal{L} = \mathcal{L}^{\rm free} + \mathcal{L}^{\rm 4f} +
                \mathcal{L}^{\rm der}  + \mathcal{L}^{\rm em},
\end{align}
which are the free-nucleon term $\mathcal{L}^{\rm free}$
\begin{align}
  \mathcal{L}^{\rm free}=\bar\psi(i\gamma_\mu\partial^\mu-M)\psi,
\end{align}
the four-fermion point-coupling term $\mathcal{L}^{\rm 4f}$
\begin{align}\label{equ:L-4f}
  \mathcal{L}^{\rm 4f}=&-\frac{1}{2}\alpha_{S}(\bar\psi\psi)(\bar\psi\psi)
  -\frac{1}{2}\alpha_{tS}(\bar\psi\vec\tau\psi)
                          (\bar\psi\vec\tau\psi) \nonumber\\
  &-\frac{1}{2}\alpha_{V}(\bar\psi\gamma_\mu\psi)
                         (\bar\psi\gamma^\mu\psi)
  -\frac{1}{2}\alpha_{tV}(\bar\psi\gamma_\mu\vec\tau\psi)
                          (\bar\psi\gamma^\mu\vec\tau\psi) \nonumber\\
  &-\frac{1}{2}\alpha_{T}(\bar\psi\sigma_{\mu\nu}\psi)
                         (\bar\psi\sigma^{\mu\nu}\psi)
  -\frac{1}{2}\alpha_{tT}(\bar\psi\sigma_{\mu\nu}\vec\tau\psi)
                          (\bar\psi\sigma^{\mu\nu}\vec\tau\psi) \nonumber\\
  &-\frac{1}{2}\alpha_{PS}(\bar\psi\gamma_5\psi)
                          (\bar\psi\gamma_5\psi)
  -\frac{1}{2}\alpha_{tPS}(\bar\psi\gamma_5\vec{\tau}\psi)
                           (\bar\psi\gamma_5\vec{\tau}\psi) \nonumber\\
  &-\frac{1}{2}\alpha_{PV}(\bar\psi\gamma_5\gamma_\mu\psi)
                          (\bar\psi\gamma_5\gamma^\mu\psi)
  -\frac{1}{2}\alpha_{tPV}(\bar\psi\gamma_5\gamma_\mu\vec{\tau}\psi)
                           (\bar\psi\gamma_5\gamma^\mu\vec{\tau}\psi),
\end{align}
the derivative term $\mathcal{L}^{\rm der}$
\begin{align}\label{equ:L-der}
  \mathcal{L}^{\rm der}=-\frac{1}{2}\delta_S\partial_\mu(\bar\psi\psi)
                       \partial^\mu(\bar\psi\psi),
\end{align}
and the electromagnetic interaction term $\mathcal{L}^{\rm em}$
\begin{align}
  \mathcal{L}^{\rm em}=-e\frac{1-\tau_3}{2}\bar\psi\gamma_\mu\psi A^\mu
                        -\frac{1}{4}F_{\mu\nu}F^{\mu\nu}.
\end{align}
Here, $M$ is the nucleon mass, $e$ is the charge unit, $A_\mu$ and $F_{\mu\nu}$ are respectively the four-vector potential and strength tensor of the electromagnetic field.
The subscripts $S$, $V$, $T$, $PS$, and $PV$ stand for the scalar, vector, tensor, pseudo-scalar, and pseudo-vector couplings, respectively.
The subscript ``$t$'' refers to the corresponding isovector channel.
The derivative term is considered only in the isoscalar-scalar channel, but it is adequate to simulate the finite-range effects of the effective interaction \cite{Niksic_PRC78_034318}.
Higher-order terms or density-dependent coupling constants are usually implemented to take into account the in-medium many-body correlations.
In this work, we assume density-dependent coupling constants in the Lagrangian density.

The Hamiltonian of the system can be derived by the Legendre transformation. With the no-sea approximation, the nucleon field operator is expanded on the basis of annihilation and creation operators $\{c_\alpha,c^\dag_\alpha\}$ defined by a complete set of Dirac spinors $\{\varphi_\alpha({\bm r})\}$,
\begin{align}
  \psi(\bm r) = \sum_\alpha \varphi_\alpha(\bm r) c_\alpha,
  \qquad\qquad
  \psi^\dag(\bm r)=\sum_\alpha\varphi_\alpha^\dag(\bm r)c^\dag_\alpha.
\end{align}
In the Hartree-Fock approximation, the ground-state wave function $|\Phi_0\rangle$ is approximated by a Slater determinant,
\begin{align}
  |\Phi_0\rangle=\prod^A_{\alpha=1}c_\alpha^\dag|-\rangle,
\end{align}
in which $A$ is the number of nucleons and $|-\rangle$ represents the vacuum. The energy functional of the nuclear system is the expectation value of the Hamiltonian $H$ in the ground-state Slater determinant,
\begin{align}
  E_{\rm CDF} = \langle\Phi_0|H|\Phi_0\rangle
    = E_{\rm kin} + E_{\rm 4f} + E_{\rm der} + E_{\rm em},
\end{align}
which contains four parts, i.e., the kinetic $E_{\rm kin}$, four-fermion interaction $E_{\rm 4f}$, derivative $E_{\rm der}$, and electromagnetic $E_{\rm em}$ parts.

The kinetic energy part can be readily written as,
\begin{align}
  E_{\rm kin} = \int d^3\bm r\sum_\alpha
      \bar\varphi_\alpha(\bm r)
      (-i\bm\gamma\cdot\bm\nabla+M)
      \varphi_\alpha(\bm r).
\end{align}
For the four-fermion interaction part, both direct (Hartree) $E_{\rm H}$ and exchange (Fock) $E_{\rm F}$ terms are considered,
\begin{align}
  E_{\rm H} =& \frac{1}{2} \int d^3\bm r \sum_{i;\alpha\beta} \alpha_i^{\rm HF}
      \left[\bar\varphi_{\alpha}(\bm r)(\mathcal{O}\Gamma)_i\varphi_{\alpha}(\bm r)\right]
      \left[\bar\varphi_{\beta}(\bm r)(\mathcal{O}\Gamma)^i\varphi_{\beta}(\bm r)\right], \\
  E_{\rm F} =& -\frac{1}{2} \int d^3\bm r\sum_{i;\alpha\beta} \alpha_i^{\rm HF}
      \left[\bar\varphi_{\alpha}(\bm r)(\mathcal{O}\Gamma)_i\varphi_{\beta}(\bm r)\right]
      \left[\bar\varphi_{\beta}(\bm r)(\mathcal{O}\Gamma)^i\varphi_{\alpha}(\bm r)\right].
\end{align}
Here, the superscript of $\alpha_i^{\rm HF}$ is introduced to denote the coupling constants adopted under the Hartree-Fock approximation, and the index $i$ should run over all possible coupling channels.
The exchange terms can be expressed as the superposition of the direct terms with the Fierz transformation \cite{Greiner__b,Sulaksono2003Ann.Phys.36},
\begin{align}
  [\bar\varphi_\alpha(\mathcal{O}\Gamma)_i\varphi_\beta]
[\bar\varphi_\beta(\mathcal{O}\Gamma)^i\varphi_\alpha]
=\sum_j\Lambda_{ij}
[\bar\varphi_\alpha(\mathcal{O}\Gamma)_j\varphi_\alpha]
[\bar\varphi_\beta(\mathcal{O}\Gamma)^j\varphi_\beta]
\end{align}
with $\Lambda$ being the Fierz transformation matrix.
As a result, the four-fermion part $E_{\rm 4f}$ can be written as
\begin{align}
  E_{\rm 4f} = E_{\rm H}+E_{\rm F}
  =\frac{1}{2} \int d\bm r \sum_{i;\alpha\beta} \alpha_i
   \left[\bar\varphi_{\alpha}(\bm r)(\mathcal{O}\Gamma)_i\varphi_{\alpha}(\bm r)\right]
   \left[\bar\varphi_{\beta}(\bm r)(\mathcal{O}\Gamma)^i\varphi_{\beta}(\bm r)\right],
\end{align}
where $\alpha_i$ is defined as
\begin{align}
  \alpha_i = \sum_jC_{ij}\alpha_j^{\rm HF}, \qquad
  i,j=\{S,tS,V,tV,T,tT,PS,tPS,PV,tPV\}
\end{align}
with the matrix $C$
\begin{align}
  C = 1-\Lambda^T
=\frac{1}{16}
\left(\begin{array}{rrrrrrrrrr}
 14 & -6 & -8 & -24 & -24 & -72 & -2 & -6 &  8 &  24 \\
 -2 & 18 & -8 &   8 & -24 &  24 & -2 &  2 &  8 &  -8 \\
 -2 & -6 & 20 &  12 &   0 &   0 &  2 &  6 &  4 &  12 \\
 -2 &  2 &  4 &  12 &   0 &   0 &  2 & -2 &  4 &  -4 \\
 -1 & -3 &  0 &   0 &  20 &  12 & -1 & -3 &  0 &   0 \\
 -1 &  1 &  0 &   0 &   4 &  12 & -1 &  1 &  0 &   0 \\
 -2 & -6 &  8 &  24 & -24 & -72 & 14 & -6 & -8 & -24 \\
 -2 &  2 &  8 &  -8 & -24 &  24 & -2 & 18 & -8 &   8 \\
  2 &  6 &  4 &  12 &   0 &   0 & -2 & -6 & 20 &  12 \\
  2 & -2 &  4 &  -4 &   0 &   0 & -2 &  2 &  4 &  12 \\
\end{array}\right).
\end{align}
Note that the rank of the matrix $C$ is 5, so once the coupling constants  $\alpha_i$ in five channels are known, one can immediately obtain the ones in the remain channels.
In the present work, the coupling constants in the channels of $S$, $tS$, $V$, $tV$, and $T$ are regarded as the free parameters to be determined, so the ones in other channels read
\begin{subequations}\label{equ:coupling-relation}
  \begin{align}
  \alpha_{ tT} &= \frac{1}{18}(- \alpha_S+3\alpha_{tS}+ 2\alpha_V- 6\alpha_{tV}+ 6\alpha_T) , \label{equ:coupling-relation1} \\
  \alpha_{ PS} &= \frac{1}{ 3}(- \alpha_S-6\alpha_{tS}- 4\alpha_V+12\alpha_{tV}-12\alpha_T) ,\\
  \alpha_{tPS} &= \frac{1}{ 9}(-4\alpha_S+3\alpha_{tS}+ 8\alpha_V-24\alpha_{tV}-12\alpha_T) , \label{equ:coupling-relation3} \\
  \alpha_{ PV} &= \frac{1}{ 3}( 2\alpha_S+3\alpha_{tS}+ 2\alpha_V+ 3\alpha_{tV}+ 6\alpha_T) ,\\
  \alpha_{tPV} &= \frac{1}{ 9}( 2\alpha_S+3\alpha_{tS}+ 5\alpha_V- 6\alpha_{tV}+ 6\alpha_T) . \label{equ:coupling-relation2}
  \end{align}
\end{subequations}
With the redefined coupling constants $\alpha_i$, the exchange terms are now treated in a similar way to the direct ones.

The derivative $E_{\rm der}$ and electromagnetic $E_{\rm em}$ parts can be respectively written as
\begin{align}
  E_{\rm der} = -\frac{1}{2} \int d^3\bm{r} \sum_{\alpha\beta}\delta_S
                \bm\nabla\left[\bar\varphi_\alpha(\bm{r})\varphi_\alpha(\bm{r})\right]
                \bm\nabla\left[\bar\varphi_\beta(\bm{r})\varphi_\beta(\bm{r})\right],
\end{align}
and
\begin{align}
  E_{\rm em} = \frac{1}{2}\iint d^3\bm r d^3\bm r'
               \sum_{\alpha\beta} \frac{e^2}{4\pi}
               \left[\bar\varphi_{\alpha}(\bm r)\gamma_\mu\frac{1-\tau_3}{2}\varphi_{\alpha}(\bm r)\right]
               \frac{1}{|\bm r-\bm r'|}
               \left[\bar\varphi_{\beta}(\bm r')\gamma^\mu\frac{1-\tau_3}{2}\varphi_{\beta}(\bm r')\right].
\end{align}
Here, for simplicity reason, the exchange terms are neglected.

Pairing correlations are essential for open-shell nuclei, so the pairing energy, which depends on the pairing density, should also be included in the energy density functional,
\begin{align}
  E_{\rm pair} = \frac{1}{2}\rm{Tr}[\Delta\kappa].
\end{align}
Here, $\kappa$ is the pairing density (see below), and $\Delta$ is the pairing field, whose matrix elements can be written as
\begin{align}
  \Delta_{a b}=\frac{1}{2} \sum_{cd}\left\langle a b\left|V^{p p}\right| c d\right\rangle \kappa_{c d},
\end{align}
with $V^{pp}$ being the pairing force. In this work, a separable form of the finite-range interaction Gogny D1S is adopted for the pairing force \cite{Tian_PLB676_44},
\begin{align}\label{equ:separable-pairing}
  V^{p p}\left(\boldsymbol{r}_{1}, \boldsymbol{r}_{2}, \boldsymbol{r}_{1}^{\prime}, \boldsymbol{r}_{2}^{\prime}\right)=-G \delta\left(\boldsymbol{R}-\boldsymbol{R}^{\prime}\right) P(\boldsymbol{r}) P\left(\boldsymbol{r}^{\prime}\right) \frac{1}{2}\left(1-P^{\sigma}\right),
\end{align}
in which $\bm R=(\bm r_1+\bm r_2)/2$, $\bm r=\bm r_1 -\bm r_2$, and $P(\bm r)$ has a Gaussian expression
\begin{align}
  P(\boldsymbol{r})=\frac{1}{\left(4 \pi a^{2}\right)^{3 / 2}} e^{-r^{2} / 4 a^{2}}.
\end{align}
The constants $G$ and $a$ are two parameters for the pairing density functional.

The variation of the energy density functional results in the relativistic Hartree-Fock-Bogoliubov (RHFB) equation,
\begin{align}\label{equ:RHFB}
  \left(\begin{array}{cc}
    \hat{h}-\lambda & \hat{\Delta} \\
    -\hat{\Delta}^{*} & -\hat{h}^{*}+\lambda
    \end{array}\right)\left(\begin{array}{l}
    U_\alpha \\
    V_\alpha
    \end{array}\right)=E_\alpha\left(\begin{array}{l}
    U_\alpha \\
    V_\alpha
    \end{array}\right),
\end{align}
where $\hat{h}$ is the single-particle Dirac Hamiltonian, $\hat{\Delta}$ is the pairing field, $\lambda$ is the Fermi energy, $E_\alpha$ is the quasiparticle energy, and $U_\alpha$ and $V_\alpha$ are the quasiparticle wavefunctions.
The RHFB equation gives a unified and self-consistent treatment of the mean field and the pairing field \cite{Kucharek_ZPA-HaN339_23,Gonzalez-Llarena_PLB379_13,Meng_NPA635_3}.

Here, the single-particle Hamiltonian $\hat{h}$ reads
\begin{align}
  \hat{h}=\bm\alpha \cdot \bm p +\beta \left(M+S\right) + V - \bm\alpha \cdot \bm V
          -i\beta\bm\alpha\cdot\bm{T}^0
\end{align}
which is coupled with the scalar, vector, and tensor potentials
\begin{align}
  S &= \alpha_S\rho_S + \alpha_{tS}\tau_3\rho_{tS}
         + \delta_S\Delta\rho_S, \\
  V^0 &= \alpha_V \rho_V + \alpha_{tV}\tau_3\rho_{tV}
         + e\frac{1-\tau_3}{2}A^0,   \\
  \bm{V} &= \alpha_V \bm{j}_V + \alpha_{tV}\tau_3\bm{j}_{tV}
         + e\frac{1-\tau_3}{2}\bm{A},   \\
  \bm{T}^0 &= 2\alpha_T \bm{j}^0_T + 2\alpha_{tV}\tau_3\bm{j}^0_{tT},
\end{align}
via various densities and currents
\begin{align}
\rho_S    &=\sum_{\alpha>0}\bar{V}_\alpha V_\alpha,\quad &
\rho_{tS} &=\sum_{\alpha>0}\bar{V}_\alpha\tau_3 V_\alpha, \\
\rho_V    &=\sum_{\alpha>0}\bar{V}_\alpha\gamma^0V_\alpha,\quad &
\rho_{tV} &=\sum_{\alpha>0}\bar{V}_\alpha\gamma^0\tau_3 V_\alpha, \\
\bm{j}_V    &=\sum_{\alpha>0}\bar{V}_\alpha \bm\gamma V_\alpha,\quad &
\bm{j}_{tV} &=\sum_{\alpha>0}\bar{V}_\alpha \bm\gamma \tau_3 V_\alpha, \label{equ:currents-V} \\
\bm{j}^0_T    &=\sum_{\alpha>0}\bar{V}_\alpha i\gamma^0\bm{\gamma} V_\alpha, \quad &
\bm{j}^0_{tT} &=\sum_{\alpha>0}\bar{V}_\alpha i\gamma^0\bm{\gamma} \tau_3 V_\alpha. \label{equ:currents-T}
\end{align}
In addition, the pairing density reads $\kappa=\sum\limits_{\alpha>0} V^\ast_\alpha U^T_\alpha$.
According to the no-sea approximation, the sum of $\alpha$ runs over all the quasiparticle states with positive energies in the Fermi sea.
For a system with time-reversal invariance, the spatial components of the currents $\bm j_V$, $\bm j_{tV}$ in Eqs. (\ref{equ:currents-V}) and (\ref{equ:currents-T}) vanish.
One should also note that the $\sigma^{ij}$ ($i,j =1, 2,$ and $3$) component of the tensor channel, the pseudo-scalar and pseudo-vector channels give no contribution in infinite nuclear matter and the ground state of finite nuclei in the present framework.

The center-of-mass (c.m.) correction energy $E_{\rm c.m.}$ should be taken into account for finite nuclei due to the breaking of the translational symmetry. It is estimated with a microscopic method \cite{Bender_EA7_467a,Long_PRC69_034319,Zhao_CPL26_112102}
\begin{align}
  E_{\rm c.m.} = -\frac{\langle\bm P^2_{\rm c.m.}\rangle}{2MA},
\end{align}
where $\bm P_{\rm c.m.}=\sum_i^A\bm p_i$ is the total momentum in the c.m. frame.

\section{Numerical Details}\label{sec:numerical}
The localized RHFB equation (\ref{equ:RHFB}) is solved in the space of harmonic oscillator wave functions \cite{Niksic_CPC185_1808}.
The harmonic oscillator basis in the spherical and Cartesian frame are respectively used for spherical and deformed calculations, and the oscillator frequencies are taken as $\hbar\omega_0=70A^{-1/3}$ MeV.
In the present work, it includes 30 major shells for the spherical calculations, and 24 major shells for the deformed cases.
It has been checked that for the spherical cases, by increasing the major shells from $N_f=30$ to $N_f=32$, the binding energy, the charge radius, and the neutron skin thickness of $^{208}$Pb change by $0.0006\%$, $0.0004\%$, and $0.05\%$, respectively.
For the deformed cases, the variation of the binding energy of $^{240}$U is within $0.007\%$ from $N_f=24$ to $N_f=26$.

In this work, the least-square fit with the penalty function
\begin{align}\label{equ:chi2}
  \chi^2(\bm p) = \sum^N_i\left(
           \frac{\mathcal{O}^{\rm exp.}_i-\mathcal{O}^{\rm cal.}_i(\bm p)}{\Delta\mathcal{O}_i}
           \right)^2,
\end{align}
is employed to determine the parameters in the density functional. Here, the vector $\bm p$ represents the ensemble of the parameters, $\mathcal{O}^{\rm exp.}_i$ are the experimental data or pseudo-data from the \emph{ab initio} calculations, $\mathcal{O}^{\rm cal.}_i(\bm p)$ are the predicted values from the covariant DFT, and $\Delta\mathcal{O}_i$ denotes the adopted weights for the selected observables.
The multi-parameter fitting is carried out by the \emph{lmfit} package in python with the Levenberg-Marquardt algorithm \cite{newville_matthew_2014_11813}.

\section{Strategy of the parametrization}\label{sec:strategy}

By taking into account the localized exchange terms with the Fierz transformation, there are totally 11 coupling constants in the Lagrangian density (\ref{equ:Lagrangian}) to be determined, i.e.,  $\alpha_S$, $\alpha_V$, $\alpha_{tS}$, $\alpha_{tV}$, $\alpha_T$, $\alpha_{tT}$, $\alpha_{PS}$, $\alpha_{tPS}$, $\alpha_{PV}$, $\alpha_{tPV}$, and $\delta_S$.
However, due to the relations given in Eqs. (\ref{equ:coupling-relation1}) - (\ref{equ:coupling-relation2}), only 6 of them are free parameters, which are $\alpha_S$, $\alpha_V$, $\alpha_{tS}$, $\alpha_{tV}$, $\alpha_T$, and $\delta_S$.
The first four coupling constants $\alpha_S$, $\alpha_V$, $\alpha_{tS}$, and $\alpha_{tV}$ are assumed to be density-dependent,
\begin{align}\label{equ:coupling-ddf}
  \alpha_i(\rho) = \alpha_i(\rho_{\rm sat.})f_i(x), \quad\text{for}\quad i = S, V, tS, \text{ and } tV,
\end{align}
where $\rho_{\rm sat.}$ is the saturation density and $x=\rho/\rho_{\rm sat.}$. The density-dependent function $f_i(x)$ is taken as the following ansatz \cite{Typel_NPA656_331},
\begin{align}\label{equ:dd-function}
  f_i(x) = a_i\frac{1+b_i(x+d_i)^2}{1+c_i(x+d_i)^2}.
\end{align}
It should satisfy $f_i(1)=1$ by definition, and another constraint $f''_i(0)=0$ is imposed as in Ref. \cite{Typel_NPA656_331} to further reduce the number of parameters.
Therefore, three free parameters are needed for the density dependence in each channel. Here, we take $\alpha_i(\rho_{\rm sat.})$, $a_{i}$, and $d_{i}$ as the free parameters.
In addition, the coupling constants $\alpha_T$ and $\delta_S$ are assumed to be density independent.

In the following, the parameters involved in the energy density functional will be determined by using the bulk properties of infinite nuclear matter and the ground state of finite nuclei step by step.

\subsection{Infinite nuclear matter}

First, we determine the coupling constants in the isoscalar ($S$ and $V$) channels as well as their density dependence, i.e., $\alpha_S(\rho)$ and $\alpha_V(\rho)$.
The 6 parameters are determined by the properties of the symmetric nuclear matter, as listed in Table \ref{tab:obs-sym}, which include the saturation density, energy per nucleon at saturation, Dirac mass and compression modulus at the saturation density, as well as the energies per nucleon at two densities respectively below and above the saturation density~\cite{Akmal_PRC58_1804}.
These properties are solely determined by the isoscalar channels.

\begin{table*}[!htbp]
  \caption{The properties of symmetric nuclear matter used to determine the coupling constants in the isoscalar channels. They include the saturation density $\rho_{\rm sat.}$, energy per nucleon at saturation $E/A|_{\rho_{\rm sat.}}$, Dirac Mass $M^*_D$ and the compression modulus $K$ at the saturation density, as well as the energies per nucleon $E/A$ at two additional densities $\rho=0.04 \text{ fm}^{-3} \text{ and }0.56 \text{ fm}^{-3}$, which are taken from Ref.~\cite{Akmal_PRC58_1804}.
  }\label{tab:obs-sym}
  \begin{tabular}{cc}
    \hline\hline
    Nuclear matter properties & Values \\ \hline
    $\rho_{\rm sat.}$ (fm$^{-3}$)
      & 0.152, 0.154, 0.156, 0.158, 0.160 \\
    $E/A|_{\rho_{\rm sat.}}$  (MeV)
      & -16.00, -16.02, -16.04, -16.06, -16.08, -16.10, -16.12, -16.14, -16.16 \\
    $M^*_D/M$
      & 0.58, 0.70, 0.80 \\
    $K$ (MeV)
      & 230 \\
    $E/A|_{\rho=0.04\text{ fm}^{-3}}$ (MeV)
      & -6.48 \\
    $E/A|_{\rho=0.56\text{ fm}^{-3}}$ (MeV)
      & 34.39 \\
    \hline\hline
  \end{tabular}
\end{table*}

The properties of the symmetric nuclear matter are known only in empirical regions.
For the saturation density $\rho_{\rm sat.}$ and the corresponding energy per nucleon $E/A|_{\rho_{\rm sat.}}$, 5 and 9 values in the empirical regions are respectively adopted for the fitting, since they have pronounced influences on the properties of finite nuclei.
The Dirac mass $M_D^*$ is closely related to the spin-orbit energy splittings in finite nuclei.
It is usually known to be around $0.58M$ \cite{Niksic_PRC78_034318,Zhao_PRC82_054319}.
For the present density functional, however, a larger Dirac mass becomes possible because the tensor couplings also contribute to the spin-orbit potential. Therefore, three values $0.58$, $0.70$, and $0.80$ are chosen for $M_D^*/M$ in the fitting.
The compression modulus $K$ is taken as $K=230$ MeV, which is consistent with the notion that the experimental excitation energy of the isoscalar giant monopole resonances could be reproduced with this incompressibility \cite{Garg_PiPaNP101_55}.
In addition, the equation of state (EOS) at two densities $\rho=0.04$ fm$^{-3}$ and $\rho=0.56$ fm$^{-3}$ obtained by the \emph{ab initio} variational calculations of Akmal, Pandharipande, and Ravenhall (hereafter APR) \cite{Akmal_PRC58_1804} is used to further constrain the density dependence of $\alpha_S(\rho)$ and $\alpha_V(\rho)$.

\begin{figure}[!htbp]
  \centering
  \includegraphics[width=0.6\textwidth]{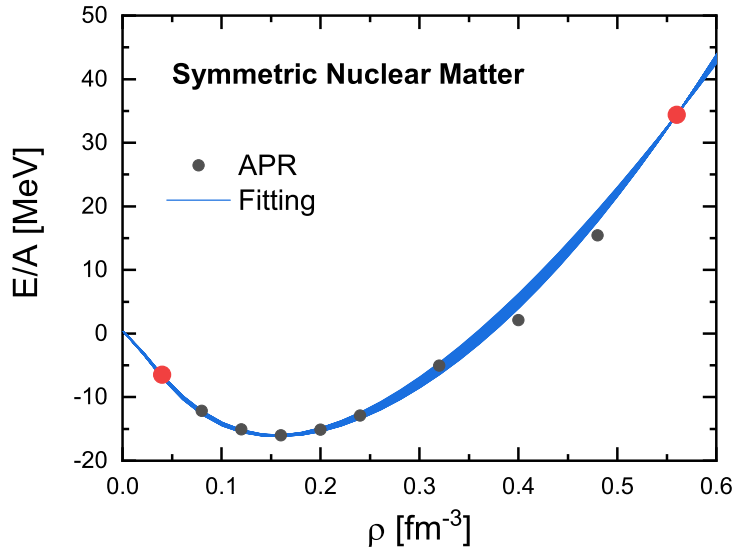}
  \caption{(Color online).
  The equation of state for symmetric nuclear matter as a function of the baryon density calculated with the 135 optimized sets of the coupling constants $\alpha_S(\rho)$ and $\alpha_V(\rho)$ in comparison with the predictions by the \emph{ab initio} variational calculations (APR) \cite{Akmal_PRC58_1804} (solid circles). The two points at the density 0.04 fm$^{-3}$ and 0.56 fm$^{-3}$ shown with larger filled circles have been used in the determination of the coupling constants.
  }
  \label{fig:EOS_fit}
\end{figure}

By fitting to the selected target properties of the symmetric nuclear matter, we have determined totally 135 sets of density-dependent coupling constants for $\alpha_S(\rho)$ and $\alpha_V(\rho)$, which respectively correspond to the 135 combinations for the fitting targets.
The fitting precision is very high, for instance, the relative deviations are within 0.005\% for the target saturation properties.
The target energies per nucleon at density 0.04 fm$^{-3}$ and 0.56 fm$^{-3}$ are also reproduced rather well, as seen in Fig. \ref{fig:EOS_fit}, where the calculated EOSs of symmetric nuclear matter given by the 135 sets of coupling constants are depicted.
Apart from the two fitted points in the EOS, one can see that the other points in the EOS given by the \emph{ab initio} variational calculations (APR) \cite{Akmal_PRC58_1804} are also well reproduced by the obtained coupling constants.
Therefore, one should not exclude any set of the coupling constants for $\alpha_S(\rho)$ and $\alpha_V(\rho)$ at this stage.

\begin{figure}[!htbp]
  \centering
  \includegraphics[width=0.6\textwidth]{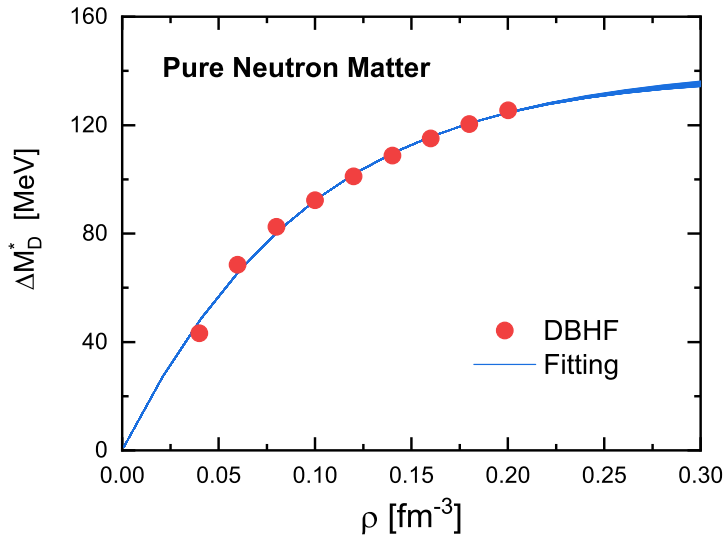}
  \caption{(Color online).
  Proton-neutron Dirac mass splitting as a function of the baryon density in pure neutron matter given by the 135 optimized sets of the coupling constants $\alpha_S(\rho)$, $\alpha_V(\rho)$, and $\alpha_{tS}(\rho)$. The solid circles represent the results from the DBHF calculations \cite{vanDalen_EPJA31_29}.}
  \label{fig:Dmass_fit}
\end{figure}

In the second step, the density-dependent coupling constants $\alpha_{tS}(\rho)$ in the isovector-scalar ($tS$) channel are determined by the Dirac mass splitting of the pure neutron matter $\Delta M^{*}_D = M^{*(p)}_D - M^{*(n)}_D$.
This quantity depends only on the $tS$ channel as long as the coupling constants in the $S$ and $V$ channels are fixed.
As there are no experimental data for the Dirac mass splittings, the results given by the Dirac-Brueckner-Hartree-Fock (DBHF) calculations \cite{vanDalen_EPJA31_29} are employed as pseudo-data for the fitting.
Note that in the DBHF calculations the scattering equations are solved with positive energy states only.
Recently, by including the negative energy states, the scattering equations are solved in the full Dirac space for both symmetric and asymmetric nuclear matter, namely the relativistic Brueckner Hartree-Fock (RBHF) theory \cite{wang2021_PRC103_054319, wang2022_ANP_}.
The values from the DBHF calculations are slightly larger than the RBHF results but qualitatively agree with the RBHF results. In the future, more results from the RBHF calculations can provide more and better guidance for the optimization of the density functional \cite{Shen2019Prog.Part.Nucl.Phys.103713}.
The density-dependent coupling constants $\alpha_{tS}(\rho)$ are determined by freezing $\alpha_S(\rho)$ and $\alpha_V(\rho)$ as determined in last step.
In Fig.~\ref{fig:Dmass_fit}, the Dirac mass splitting given by the 135 optimized sets of the coupling constants $\alpha_S(\rho)$, $\alpha_V(\rho)$, and $\alpha_{tS}(\rho)$ are depicted as a function of the nucleon density. One can see that the Dirac mass splittings can be fitted quite nicely for all the 135 sets of the coupling constants.

In the third step, the coupling constant at the saturation density in the isovector-vector ($tV$) channel $\alpha_{tV}(\rho_{\rm sat.})$ is solely determined by the symmetry energy $J$.
In the present work, the symmetry energy $J$ is taken as $J=33$ MeV, which is also consistent with results of the \emph{ab initio} calculations~\cite{Akmal_PRC58_1804,Baldo_NPA736_241,wang2022_ANP_}.
Freezing $\alpha_S(\rho)$, $\alpha_V(\rho)$, and $\alpha_{tS}(\rho)$ at the values found in steps 1 and 2, the coupling constant $\alpha_{tV}(\rho_{sat.})$ is determined by reproducing $J=33$ MeV within 0.001\%.
Note that due to the relatively large uncertainties of the density-dependence of the symmetry energy, the density-dependence of $\alpha_{tV}(\rho)$ cannot be well constrained, so they are left to be determined by the observables of finite nuclei.

\subsection{Finite Nuclei}

Apart from the density-dependence of $\alpha_{tV}(\rho)$, the coupling constants $\alpha_T$ and $\delta_S$ are also optimized by fitting to the bulk properties of spherical nuclei, because they only affect the properties of finite nuclei.

\begin{figure*}[!htbp]
  \centering
  \includegraphics[width=0.6\textwidth]{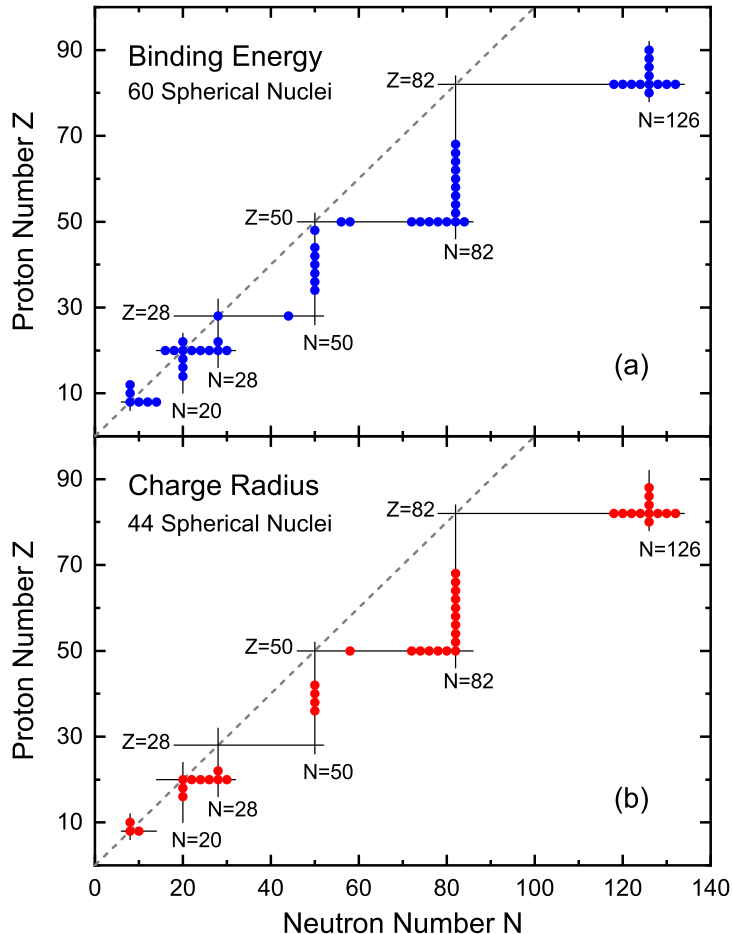}
  \caption{(Color online).
  Nuclei whose binding energies (a) and charge radii (b) are used in the fitting of the present density functionals.}
  \label{fig:chosen-nuclei}
\end{figure*}

The binding energies of 60 spherical nuclei \cite{Wang_CPC41_030003} and the charge radii of 44 ones \cite{Angeli_ADaNDT99_69} are selected as the observables to be fitted, and they are shown in Figs. \ref{fig:chosen-nuclei}(a) and \ref{fig:chosen-nuclei}(b), respectively.
The two-neutron separation energies of $^{18}$O, $^{42}$Ca, $^{50}$Ca, $^{134}$Sn, $^{210}$Pb and the two-proton separation energies of $^{18}$Ne, $^{42}$Ti, $^{50}$Ti, $^{134}$Te, $^{210}$Po are also fitted to achieve a reasonable description of the major shell gaps.
In addition, the empirical proton pairing gaps of $^{92}$Mo, $^{136}$Xe, $^{144}$Sm, and the neutron ones of $^{122}$Sn,  $^{124}$Sn, $^{208}$Pb obtained with the three-point formula are also employed to constrain the paring strength $G$ in Eq.~(\ref{equ:separable-pairing}).
The adopted weights $\Delta\mathcal{O}_i$ for the binding energies, charge radii, two-nucleon separation energies, and pairing gaps, are
 1.0 MeV, 0.01 fm, 0.07 MeV, and 0.05 MeV, respectively.
Keeping the parameters as found with the nuclear matter properties, all the remaining parameters including $a_{tV}$, $d_{tV}$, $\alpha_T$, $\delta_S$, and $G$ are fitted by minimizing the $\chi^2$ in Eq. (\ref{equ:chi2}) for the observables of finite nuclei.

\begin{figure*}[!htbp]
  \centering
  \includegraphics[width=0.7\textwidth]{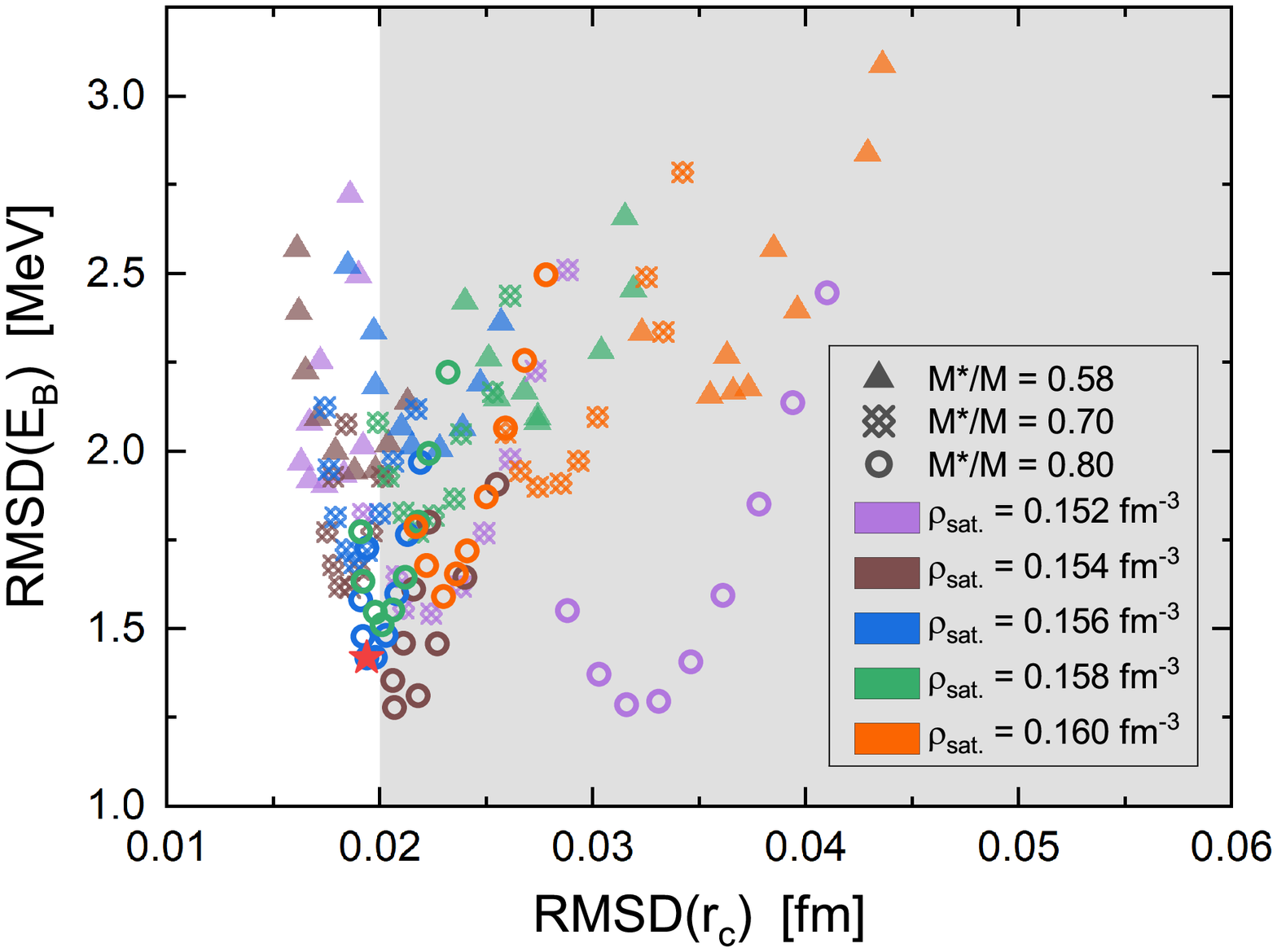}
  \caption{(Color online).
  The root-mean-square (RMS) deviations for the binding energies versus those for the charge radii given by the 135 optimized density functionals.
  The shaded area denotes the region with the $rms$ deviations for the charge radius larger than 0.02 fm.
  The symbols and colors represent different Dirac masses and saturation densities of the corresponding density functionals, respectively.
  The new density functional PCF-PK1 is represented with the star.
  }
  \label{fig:RMSD_BE-rc}
\end{figure*}

In such a way, 135 sets of the covariant density functional with localized exchange terms have been obtained.
In Fig. \ref{fig:RMSD_BE-rc}, the root-mean-square (RMS) deviations for the binding energies versus those for the charge radii given by the obtained 135 density functionals are depicted.
The functionals with the $rms$ deviations for the charge radii beyond $0.02$ fm, as shown in the shaded region, are excluded since most of the popular covariant density functionals in the market could describe the charge radii on the same level~ \cite{Long_PRC69_034319,Lalazissis_PRC71_024312,Zhao_PRC82_054319,Roca-Maza_PRC84_054309}.
For the rest of the density functionals, the one that has the least $rms$ deviation for the binding energies,
as marked with the star, is chosen to be the most optimized density functional.
Hereafter, this density functional will be called PCF-PK1, whose parameters are listed in Table \ref{tab:PCF-PK1}.

Note that not all parameters in Table \ref{tab:PCF-PK1} are independent, in fact, it includes 14 independent parameters in the mean-field channel, 10 of which are determined by the empirical saturation properties and the pseudo-data obtained from the \emph{ab initio} calculations of infinite nuclear matter~\cite{Akmal_PRC58_1804,vanDalen_EPJA31_29}.
The remaining 4 parameters ($a_{tV}$, $d_{tV}$, $\alpha_T$, and $\delta_S$),
as well as the pairing strength $G$ in the pairing channel are fitted to the masses and charge radii of finite nuclei.

\begin{table}[!htbp]
\caption{Parameters of the relativistic point-coupling functional PCF-PK1 with localized exchange terms. The nucleon mass $M$ is taken 939.0 MeV and the saturation density $\rho_{\rm sat.}$ is 0.156 fm$^{-3}$. It includes 14 independent parameters, only 4 of which ($a_{tV}$, $d_{tV}$, $\alpha_T$, and $\delta_S$) are fitted to finite nuclei, and the other 10 are derived from an adjustment to empirical saturation properties and \emph{ab initio} calculations of infinite nuclear matter~\cite{Akmal_PRC58_1804,vanDalen_EPJA31_29}. The strength and width parameters of the separable pairing force are also listed in the last row.}
\label{tab:PCF-PK1}
\begin{tabular}{cccccc}
  \hline\hline
   $i$  & $\alpha_i(\rho_{\rm sat.})$ [fm$^2$] &  $a_i$ & $b_i$ & $c_i$ & $d_i$     \\
   \hline
   S  & -6.315494 &  1.425216  & 0.3185493  & 0.592876  & 0.7498207 \\
   V  & 3.533254  &  0.8075938 & 0.0605804 & 0.03692602 & 3.004506  \\
   tS & -1.980515 &  2.328043  & 0.06178842 & 0.5704278  & 0.7644323 \\
   tV & 2.975891  &  2.546886  & 0.3459254  & 1.611987   & 0.4547352 \\
   \hline
   $\alpha_T$ [fm$^2$]  & 3.373807 & $\delta_S$ [fm$^4$]  & -0.665808 & & \\
   \hline
   $G$ [MeV$\cdot$fm$^3$]  & 657.5419 & $a$ [fm] & 0.644  & & \\
  \hline\hline
 \end{tabular}
\end{table}

In this work, calculations with the density functional PC-PK1 \cite{Zhao_PRC82_054319}, DD-PC1 \cite{Niksic_PRC78_034318}, and DD-ME$\delta$ \cite{Roca-Maza_PRC84_054309} are also performed for comparison, in which the separable pairing force with $G=728$ MeV$\cdot$fm$^{-3}$ is used.
Note that since the pairing force is not the ones used in the determination of the three density functionals, we obtain slightly different results for each density functional as compared to those given in Refs. \cite{Niksic_PRC78_034318,Roca-Maza_PRC84_054309,Zhao_PRC82_054319}.

\begin{figure*}[!htbp]
  \centering
  \includegraphics[width=0.6\textwidth]{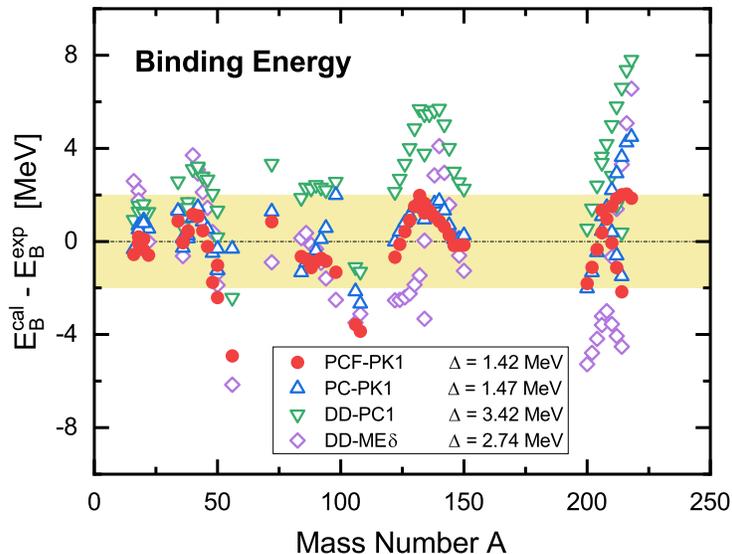}
  \caption{(Color online).
  Deviations of the theoretical binding energies obtained with PCF-PK1, PC-PK1, DD-PC1, and DD-ME$\delta$ from the experimental values \cite{Wang_CPC41_030003} for nuclei used in the fitting of PCF-PK1. The shaded region corresponds to deviations within $\pm 2$ MeV. The $rms$ deviations $\Delta=\sqrt{\sum_i^N(\mathcal{O}_i^{\rm exp.}-\mathcal{O}_i^{\rm cal.})^2/N}$ for each density functional are also listed.
  }
  \label{fig:fit-be}
\end{figure*}

\begin{figure*}[!htbp]
  \centering
  \includegraphics[width=0.6\textwidth]{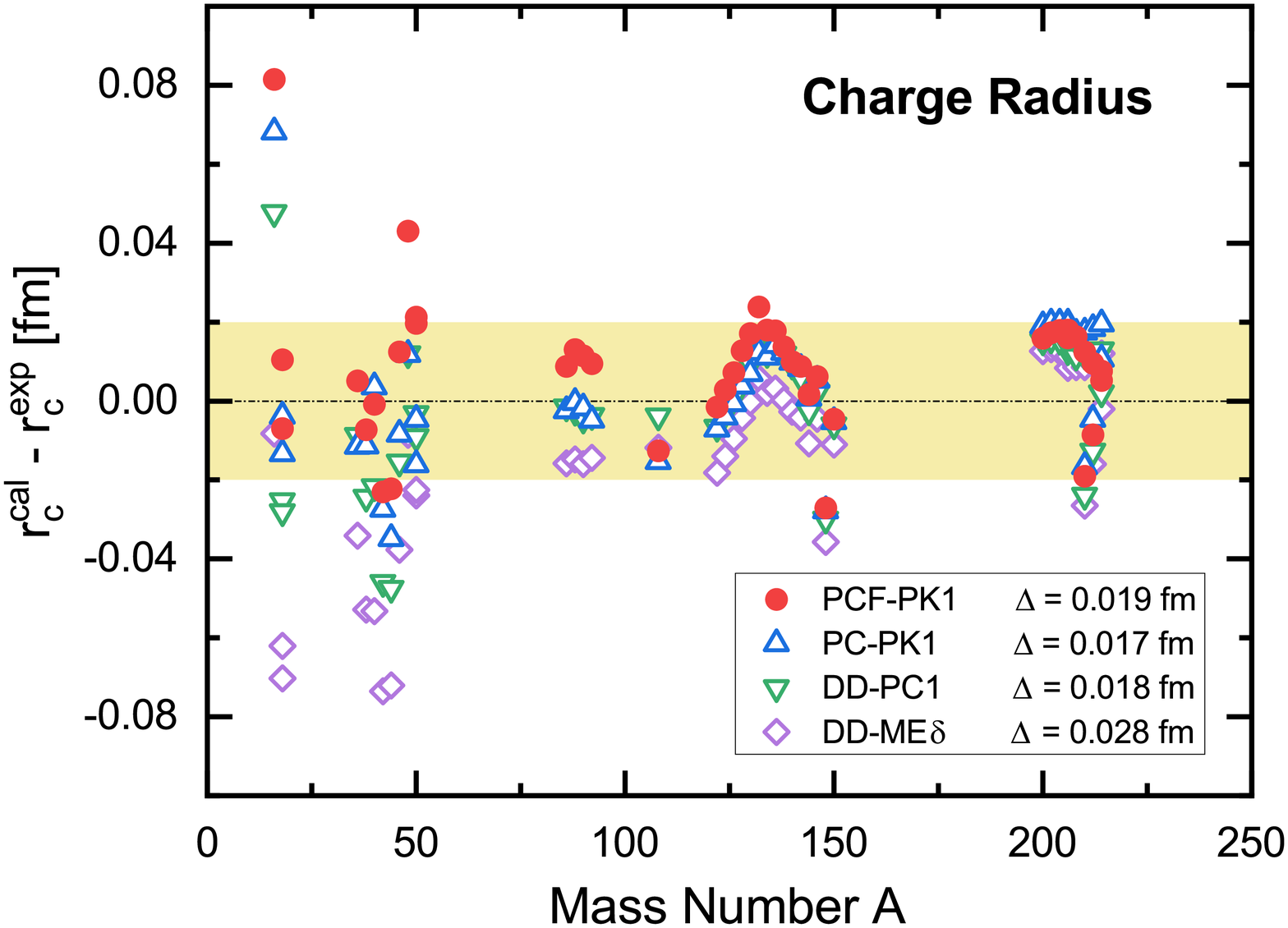}
  \caption{(Color online).
  Same as Fig. \ref{fig:fit-be}, but for charge radii. The experimental data are taken from \cite{Angeli_ADaNDT99_69}.
  }
  \label{fig:fit-rch}
\end{figure*}

Differences between the theoretical binding energies given by PCF-PK1 and the experimental ones for the selected nuclei are shown in Fig. \ref{fig:fit-be} in comparison with the results of PC-PK1, DD-PC1, and DD-ME$\delta$.
It can be seen that PCF-PK1 provides a good description for the binding energy, which is better than DD-PC1 and DD-ME$\delta$, and even slightly better than PC-PK1.
It should be kept in mind that DD-PC1 is fitted to deformed nuclei, so it seriously overestimates the binding energies of the spherical closed-shell nuclei, which leads to a large $rms$ deviation.
Apart from the binding energies, the charge radii are also calculated with PCF-PK1 for the selected nuclei, and the deviations from the experimental values are shown in Fig. \ref{fig:fit-rch} in comparison with the ones given by PC-PK1, DD-PC1, and DD-ME$\delta$.
It is found that the PCF-PK1 gives a good quantitative description for charge radii, and it is comparable to the results given by the other density functionals.
In addition, the selected pairing gaps, two-neutron separation energies, and two-proton separation energies are also well reproduced by the PCF-PK1 functional, and the corresponding $rms$ deviations are 0.15 MeV, 0.74 MeV, and 0.46 MeV, respectively.

\section{Results and Discussions}\label{sec:result}
In this section, the performance of the new density functional PCF-PK1 is illustrated with the calculations for the bulk properties of nuclear matter, the ground-state properties of spherical and deformed nuclei, and the Gamow-Teller resonances (GTR). 
The main advantages of PCF-PK1 on the effective nucleon mass for nuclear matter, the removal of spurious shell closures at $Z=58$ and $92$, and the self-consistent descriptions for the Gamow-Teller resonances will be focused.

\subsection{Effective nucleon mass of nuclear matter} 

The saturation properties of the symmetric nuclear matter including the saturation density $\rho_{\rm sat.}$, the binding energy per nucleon $E/A|_{\rho_{\rm sat.}}$, the Dirac mass $M^*_D$, the Landau mass $M^*_L$, the compression modulus $K$, the symmetry energy $J$ and its slope $L$ are calculated with the PCF-PK1 density functional, and the obtained results are listed in Table \ref{tab:saturation} in comparison with the ones given by the PC-PK1, DD-PC1, DD-ME$\delta$, and PKO2 \cite{Long_EEL82_12001} functionals, as well as the corresponding empirical values.

The main difference of PCF-PK1 from other functionals is the Dirac mass $M^*_D/M$. The value of $M^*_D/M$ given by PCF-PK1 is 0.8, and the corresponding Landau mass $M^*_L/M=[pdp/dE]_{p=p_F}/M$ is 0.85.
They are significantly larger than the ones given by other density functionals including PKO2, where the Fock terms are taken into account.
It is known that there is a strong correlation between the Dirac mass $M^*_D$ and the tensor couplings, which are responsible for the proper size of the spin-orbit splitting in finite nuclei~\cite{Furnstahl_NPA632_607}.
In the present work, due to the inclusion of the tensor couplings, the Dirac mass is enlarged to give reasonable spin-orbit splittings.
Moreover, the large Landau mass given by the PCF-PK1 implies a large single-particle level density around the Fermi energy in finite nuclei.
Note that the PKO2 provides a much larger Landau mass, but a similar Dirac mass, than the other functionals except for PCF-PK1.
This is due to the momentum dependence of the self-energies introduced by the finite-range exchange terms in PKO2, while in the present localized exchange terms, there is no momentum dependence for the self-energies.

\begin{table*}[!htbp]
\caption{The saturation properties for nuclear matter obtained by the PCF-PK1, PC-PK1, DD-PC1, DD-ME$\delta$, and PKO2 in comparison with the empirical data.
}
\label{tab:saturation}
\begin{tabular}{ccccccc}
  \hline\hline
    & Empirical & PCF-PK1 & PC-PK1 & DD-PC1 & DD-ME$\delta$ & PKO2 \\ \hline
  $\rho_{\rm sat}$ (fm$^{-3}$)
    & $0.155\pm 0.005$ \cite{Margueron_PRC97_025805}
    & 0.156  & 0.154  & 0.152  & 0.152 & 0.151 \\
  $E/A|_{\rho_{\rm sat.}}$ (MeV)
    & $-15.8\pm 0.3$ \cite{Margueron_PRC97_025805}
    & -16.10 & -16.12 & -16.06 & -16.12 & -16.00 \\
  $M^*_D/M$
    & & 0.80   & 0.59 & 0.58 & 0.61 & 0.60 \\
  $M^*_L/M$
    &    & 0.85   & 0.65   & 0.64   & 0.67   & 0.74 \\
  $K$ (MeV)
    & $230\pm 20$ \cite{Margueron_PRC97_025805}
    & 230 & 238 & 230 & 219 & 250  \\
  $J$ (MeV)
    & $31.7\pm 3.2$ \cite{Oertel_RMP89_015007}
    & 33.0   & 35.6   & 33.0   & 32.4  & 32.5  \\
  $L$ (MeV)
    & $58.7\pm 28.1$ \cite{Oertel_RMP89_015007}
    &  78.4  & 112.7  &  70.2  &  52.9   & 75.9  \\ \hline \hline
  \end{tabular}
\end{table*}

The PCF-PK1 results for the saturation properties $\rho_{\rm sat.}$, $E/A|_{\rho_{\rm sat.}}$, $K$, and $J$ are excellently consistent with the empirical values, since they are used to in the fitting procedure of the functional.
The slope $L$ of the symmetry energy at the saturation density is an important quantity that characterizes the density dependence of the symmetry energy.
The PCF-PK1 predicts the slope $L$ to be $78.4$ MeV, which is also in agreement with the corresponding empirical value \cite{Oertel_RMP89_015007}.

\begin{figure*}[!htbp]
  \centering
  \includegraphics[width=0.6\textwidth]{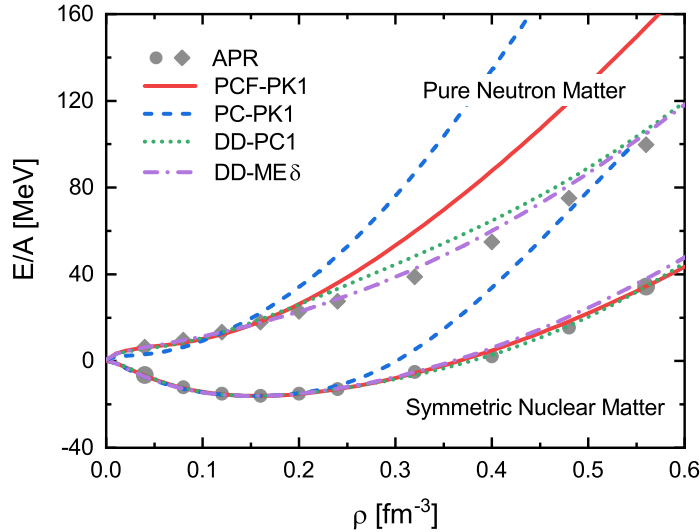}
  \caption{(Color online).
  Equation of states for the symmetric nuclear matter and the pure neutron matter calculated with the PCF-PK1, PC-PK1, DD-PC1, and DD-ME$\delta$. The \emph{ab initio} variational calculations \cite{Akmal_PRC58_1804} are shown as solid circles and diamonds for the symmetric nuclear matter and the pure neutron matter, respectively.
  }
  \label{fig:EOS}
\end{figure*}

Figure \ref{fig:EOS} depicts the EOSs of the symmetric nuclear matter and the pure neutron matter calculated with the PCF-PK1, PC-PK1, DD-PC1, and DD-ME$\delta$ in comparison with the \emph{ab initio} variational calculations (APR) \cite{Akmal_PRC58_1804}.
All density functionals give similar results that are consistent with the APR ones at lower densities, in particular, below the saturation density.
At higher densities, the DD-PC1 and DD-ME$\delta$ provide the EOSs for both symmetric nuclear matter and pure neutron matter as soft as the APR ones. Note that the EOSs of DD-ME$\delta$ is fitted to the Bruckner-Hartree-Fock (BHF) calculations \cite{Baldo_NPA736_241} which are not much different from the APR results. The DD-PC1 is not fitted to the APR EOS of pure neutron matter, but it becomes soft at higher densities due to the exponential decrease of the coupling constant of the $tV$ channel.
The PC-PK1 gives much stiffer EOSs for both symmetric nuclear matter and pure neutron matter at higher densities.
This is a common feature for most nonlinear density functionals, in which a polynomial density dependence of the coupling constants is introduced.

The PCF-PK1 gives the EOS of the symmetric nuclear matter closed to the APR one, because the APR EOS of the symmetric nuclear matter has been employed in the fitting of the PCF-PK1 functional.
The EOS of pure neutron matter given by PCF-PK1 is consistent with the APR one at the densities below 0.2 fm$^{-3}$, while it is stiffer at higher densities.
It should be noted that the existing predictions for neutron matter vary largely among different theories due to the unclear  isospin $T = 3/2$ component of the three-body force \cite{pieper2003_PRL90_252501, hammer2013_RMP85_197, zhao2016_PRC94_041302}.
Therefore, the APR EOS of pure neutron matter is not adopted in the fitting of the PCF-PK1 functional.

\subsection{Removal of the spurious shell closures at $Z=58$ and $92$}

In this part, we present single-particle levels, binding energies, two-proton shell gaps, and neutron-skin thicknesses of selected spherical isotopes and isotones obtained with the density functional PCF-PK1.

\begin{figure*}[!htbp]
  \centering
  \includegraphics[width=0.95\textwidth]{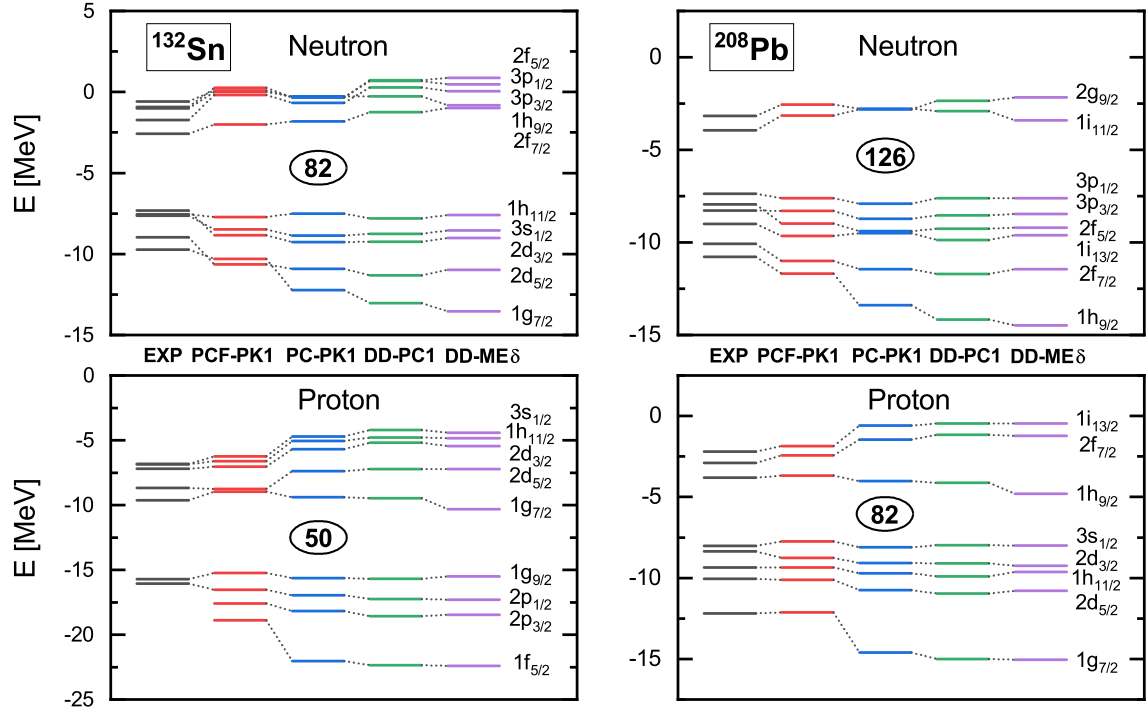}
  \caption{(Color online).
  The single-particle energies of $^{132}$Sn and $^{208}$Pb calculated by PCF-PK1, PC-PK1, DD-PC1, and DD-ME$\delta$ in comparison with the data \cite{Isakov_EPJA14_29}.
  }
  \label{fig:spe_Sn-Pb}
\end{figure*}

To examine the effect of the enhanced Dirac mass and Landau mass, the single-particle spectrum calculated by PCF-PK1 are given in Fig. \ref{fig:spe_Sn-Pb} for $^{132}$Sn and $^{208}$Pb, in comparison with those obtained by PC-PK1, DD-PC1, and DD-ME$\delta$.
The experimental values are extracted from the single-nucleon separation energies or excitation energies \cite{Isakov_EPJA14_29}.
In general, the PCF-PK1 results are in good agreement with the experimental values around the Fermi levels.
Due to the larger Landau mass, the PCF-PK1 provides higher level densities around the Fermi levels as compared with the results given by the other density functionals, and this also makes the PCF-PK1 results closer to the experimental values.
Moreover, from the energy gap between the proton $1g_{7/2}$ and $2d_{5/2}$ levels in $^{132}$Sn and the one between the proton $1h_{9/2}$ and $2f_{7/2}$ levels in $^{208}$Pb, one can clearly see the spurious shell closures at $Z=58$ and $Z=92$ predicted by all density functionals except PCF-PK1.
The removal of the spurious shell closures at $Z=58$ and $Z=92$ can be also seen from the binding energies and the two-proton shell gaps.

\begin{figure*}[!htbp]
  \centering
  \includegraphics[width=0.9\textwidth]{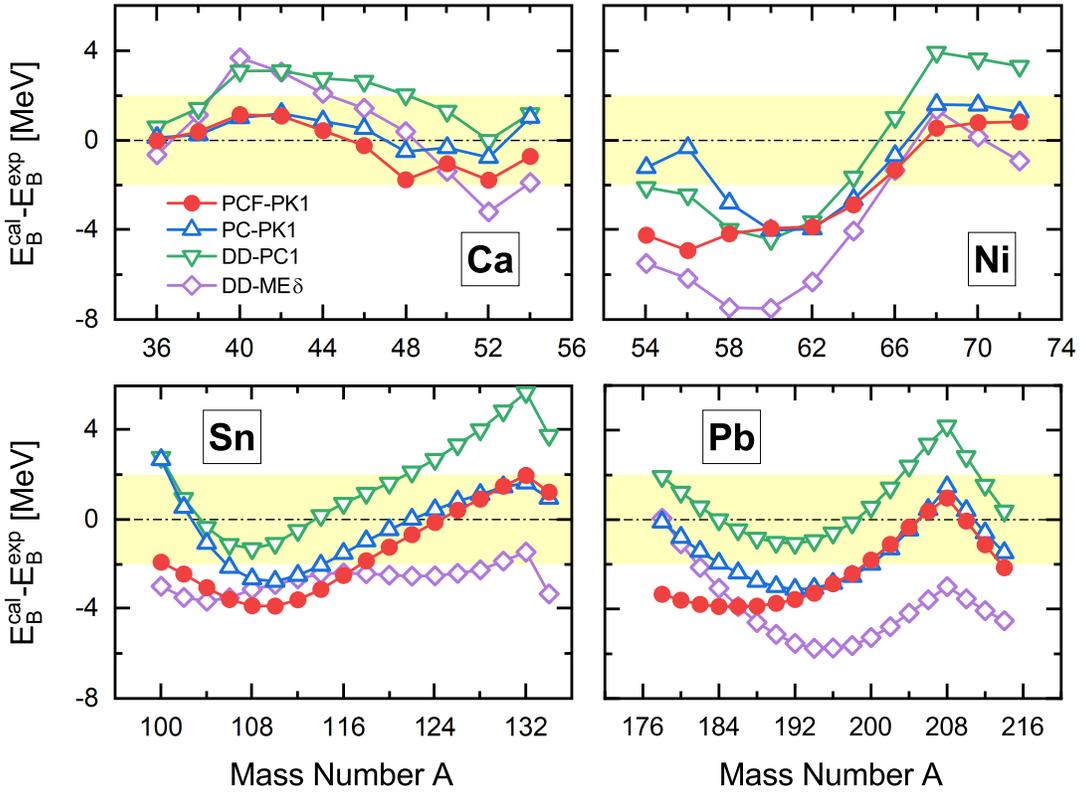}
  \caption{(Color online).
  Deviations of the binding energies of Ca, Ni, Sn, and Pb isotopes obtained by PCF-PK1, PC-PK1, DD-PC1, and DD-ME$\delta$ from the experimental values \cite{Wang_CPC41_030003}. The shaded regions represent the deviations within $\pm 2$ MeV.
  }
  \label{fig:BE_isotopes}
\end{figure*}

The binding energies of Ca, Ni, Sn, and Pb isotopes calculated with PCF-PK1 are depicted in Fig. \ref{fig:BE_isotopes} in terms of the  deviations from the experimental data \cite{Wang_CPC41_030003} in comparison with the results of PC-PK1, DD-PC1, and DD-ME$\delta$.
The PCF-PK1 results are in good agreement with the experimental data, especially for Ca isotopes, the discrepancies are less than 2 MeV.
On the neutron-deficient sides for the Ni, Sn, and Pb isotopes, the PCF-PK1 results underestimate the binding energies by up to 4 MeV.
However, for these nuclei, due to the soft potential energy surface, the dynamical correlation energy could provide more binding and, thus, improve the descriptions \cite{lu2015_PRC91_027304, yang2021_PRC104_054312}.

\begin{figure*}[!htbp]
  \centering
  \includegraphics[width=0.9\textwidth]{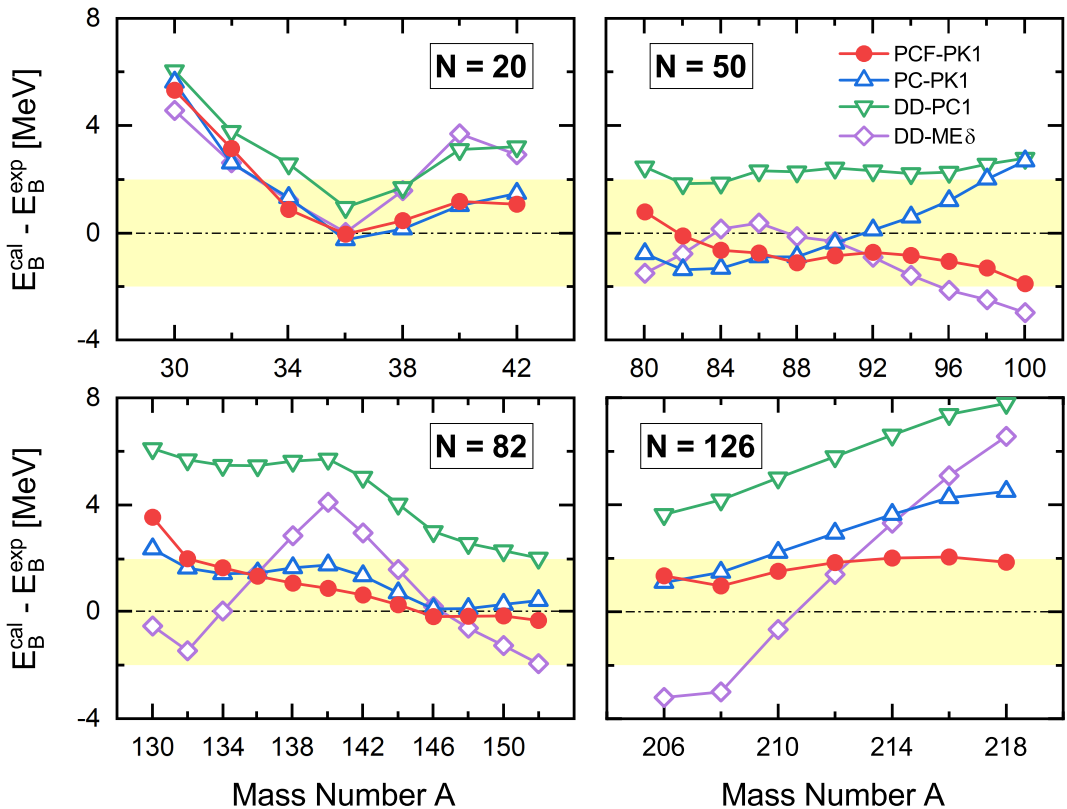}
  \caption{(Color online).
  Same as Fig. \ref{fig:BE_isotopes}, but for $N=20$, $50$, $82$, and $126$ isotones.
  }
  \label{fig:BE_isotones}
\end{figure*}

In Fig. \ref{fig:BE_isotones}, the deviations from the experimental data for the binding energies of $N=20$, $50$, $82$, and $126$ isotones calculated with the PCF-PK1 are depicted in comparison with the results of PC-PK1, DD-PC1, and DD-ME$\delta$.
The PCF-PK1 well reproduces the binding energies of these isotones within 2 MeV for most nuclei.
In particular for the $N=126$ isotones, the isospin dependence of the binding energies along the isotonic chain is improved greatly by the PCF-PK1.
For the other density functionals, with the increasing mass number, the binding energy is more and more overestimated, which leads to the so-called spurious shell closure $Z=92$ at $A=218$.

\begin{figure*}[!htbp]
  \centering
  \includegraphics[width=0.6\textwidth]{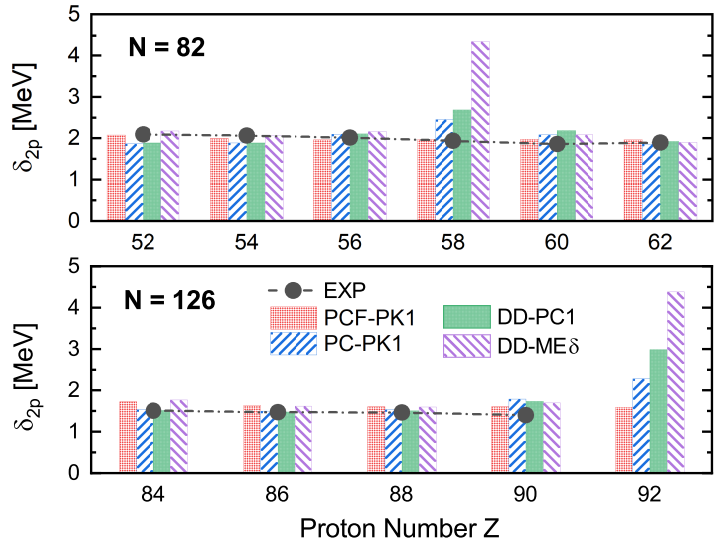}
  \caption{(Color online).
  Calculated two-proton shell gaps of $N=82$ and $126$ isotones as functions of the proton number $Z$ by the PCF-PK1, PC-PK1, DD-PC1, and DD-ME$\delta$. The experimental values extracted from Ref. \cite{Wang_CPC41_030003} are shown for comparison.
  }
  \label{fig:d2p}
\end{figure*}

To further discuss the spurious shell closures, in Fig. \ref{fig:d2p}, the two-proton shell gaps for the $N=82$ and $N=126$ isotones given by PCF-PK1 are shown in comparison with the results of PC-PK1, DD-PC1, and DD-ME$\delta$ as well as the available experimental data \cite{Wang_CPC41_030003}.
The two-proton shell gaps is defined as
\begin{align}
  \delta_{2p}(Z,N) = S_{2p}(Z,N) - S_{2p}(Z+2,N),
\end{align}
where $S_{2p}(Z,N)$ is the two-proton separation energy.
For the isotones with $N=82$, except for the nucleus with $Z=58$, all density functionals can well reproduce the experimental data.
However, only the PCF-PK1 results agree with the data at  $Z=58$, and all the other density functionals overestimate the shell gaps at $Z=58$.
Similar behaviors can also be found along the $N=126$ isotones, where all density functionals predict a substantial shell gap at $Z=92$ except PCF-PK1.
It should be noted that a recent experiment on the short-lived isotope $^{223}$Np disproves the existence of a $Z=92$ subshell closure~\cite{Sun_PLB771_303}.
Therefore, one could conclude that the spurious shell closures at $Z=58$ and $Z=92$ can be well eliminated with the new density functional PCF-PK1.
It is also worthwhile to mention that these spurious shell closures are also eliminated with the recent density functional DD-LZ1 without Fock terms, which is guided by the pseudo-spin symmetry restoration \cite{Wei_CPC44_074107}.

\begin{figure*}[!htbp]
  \centering
  \includegraphics[width=0.5\textwidth]{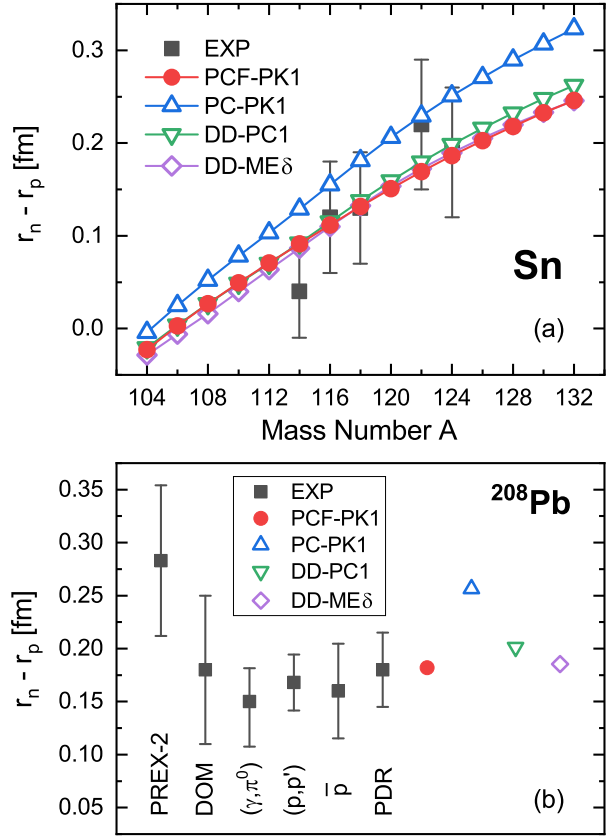}
  \caption{(Color online).
 Calculated neutron-skin thicknesses of Sn isotopes (a) and $^{208}$Pb (b) by PCF-PK1, PC-PK1, DD-PC1, and DD-ME$\delta$ in comparison with data, which are taken from Ref. \cite{Krasznahorkay_NPA567_521} for Sn isotopes, and for $^{208}$Pb, deduced from parity-violating electron scattering (PREX-2) \cite{PREXCollaboration_PRL126_172502}, dispersive optical model analysis (DOM) \cite{Pruitt_PRL125_102501}, coherent pion photoproduction ($\gamma$, $\pi^0$) \cite{CrystalBallatMAMIandA2Collaboration_PRL112_242502}, electric dipole polarizability (p,p') \cite{Roca-Maza_PRC88_024316}, antiprotonic atoms ($\bar{p}$) \cite{Klos_PRC76_014311}, and pygmy dipole resonances (PDR) \cite{LANDCollaboration_PRC76_051603}.
  }
  \label{fig:nskin}
\end{figure*}

We present in Fig. \ref{fig:nskin} the neutron-skin thicknesses of Sn isotopes and $^{208}$Pb obtained by PCF-PK1, PC-PK1, DD-PC1, and DD-ME$\delta$ in comparison with the experimental data.
For Sn isotopes, similar to other density-dependent functionals, the PCF-PK1 well describes the neutron-skin thicknesses within the experimental errors.
Note that the neutron-skin thickness is highly related to the symmetry energy, and the nonlinear density functionals, such as PC-PK1, usually provide higher symmetry energies and, thus, lead to larger neutron-skin thicknesses.

The neutron-skin thickness of $^{208}$Pb has also attracted a lot of attentions both theoretically and experimentally.
Similar to the results of Sn isotopes, the three density-dependent functionals, i.e., PCF-PK1, DD-PC1, and DD-ME$\delta$, provides similar neutron-skin thicknesses for $^{208}$Pb, while the nonlinear PC-PK1 provides a relatively larger value.
However, it should be mentioned that the corresponding experimental data have still a large uncertainty.
The data deduced from the dispersive optical model analysis \cite{Pruitt_PRL125_102501}, coherent pion photoproduction \cite{CrystalBallatMAMIandA2Collaboration_PRL112_242502}, electric dipole polarizability \cite{Roca-Maza_PRC88_024316}, antiprotonic atoms \cite{Klos_PRC76_014311}, and pygmy dipole resonances (PDR) \cite{LANDCollaboration_PRC76_051603} are in general consistent with each other, but smaller than the recent data from the parity-violating electron scattering (PREX-2).
Recent studies find that the theoretical predictions of the electric dipole polarizability that are consistent with the PREX-2 measurement systematically overestimate the corresponding values extracted from the direct measurements of the distribution of electric dipole strength \cite{piekarewicz2021_PRC104_024329,reinhard2021_PRL127_232501}.
There is yet no solution to this problem.

\subsection{Self-consistent descriptions of the Gamow-Teller resonances}

\begin{figure*}[!htbp]
  \centering
  \includegraphics[width=0.8\textwidth]{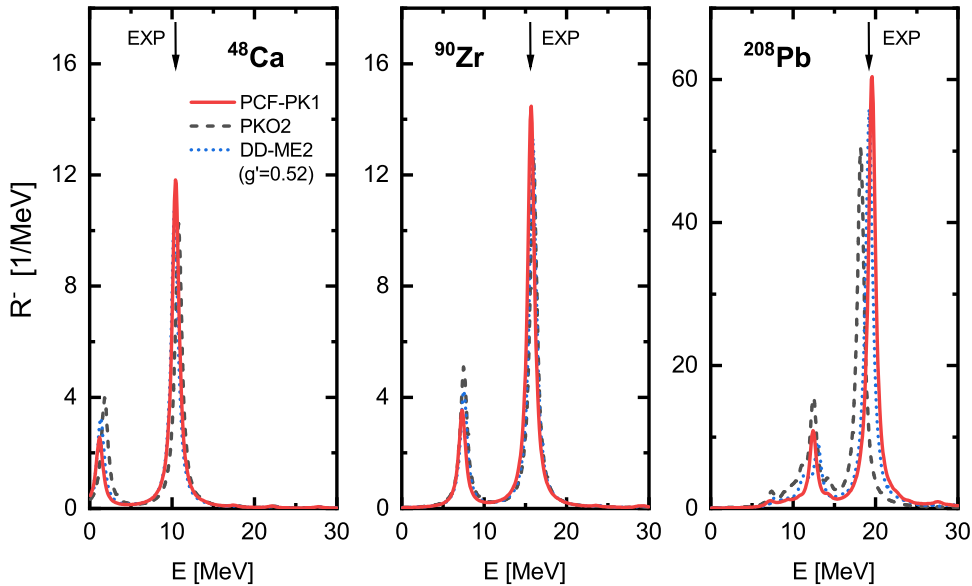}
  \caption{(Color online).
  Transition strength distributions of Gamow-Teller resonances in $^{48}$Ca, $^{90}$Zr, and $^{208}$Pb as functions of the excitation energy calculated with PCF-PK1, PKO2, and DD-ME2. A Lorentzian smearing parameter $\Gamma=1$ MeV is used. The experimental peak energies \cite{anderson1985_PRC31_1161, bainum1980_PRL44_1751, wakasa1997_PRC55_2909, horen1980_PLB95_27, akimune1995_PRC52_604} are denoted with arrows. See text for more details.
  }
  \label{fig:GTR}
\end{figure*}

Apart from the ground-state properties, the Gamow-Teller resonances are also studies with the new density functional PCF-PK1.
In particular, the transition strength distributions of Gamow-Teller resonances of $^{48}$Ca, $^{90}$Zr, and $^{208}$Pb are calculated with PCF-PK1 with the random-phase approximation (RPA), and the results are depicted in Fig. \ref{fig:GTR}.
Note that for the PCF-PK1 density functional, the $tPS$ and $tPV$ channels appear in the particle-hole (\emph{p-h}) residual interaction with the strengths determined by Eqs. (\ref{equ:coupling-relation3}) and (\ref{equ:coupling-relation2}), though they automatically vanish in the ground-state level.
For comparison, the RPA calculations based on the relativistic Hartree (RH) and RHF approaches have also been performed with DD-ME2 \cite{Lalazissis_PRC71_024312} and PKO2 \cite{Long_EEL82_12001}, respectively.
For the RH calculations with DD-ME2, the pseudo-vector pion-nucleon coupling with strength $f^2/4\pi=0.08$ is included in the \emph{p-h} residual interaction and a Landau-Migdal term with the adjustable strength $g'=0.52$ has to be employed to describe the data \cite{paar2008_PRC77_024608, Liang_PRC85_064302}.
For the RHF calculations with PKO2, however, as discussed in Ref.~\cite{Liang_PRL101_122502}, the experimental data can be well reproduced
without any additional parameters.
For the present localized RHF calculations with PCF-PK1, one can also see that the observed excitation energies for $^{48}$Ca, $^{90}$Zr, and $^{208}$Pb are reproduced nicely without any adjustments.
This demonstrates that, similar to the previous RHF and RPA calculations~\cite{Liang_PRL101_122502}, a self-consistent description for the Gamow-Teller resonances can be achieved with the present localized exchange terms in PCF-PK1.

\subsection{Ground-state properties of deformed nuclei}

\begin{figure*}[!htbp]
  \centering
  \includegraphics[width=0.9\textwidth]{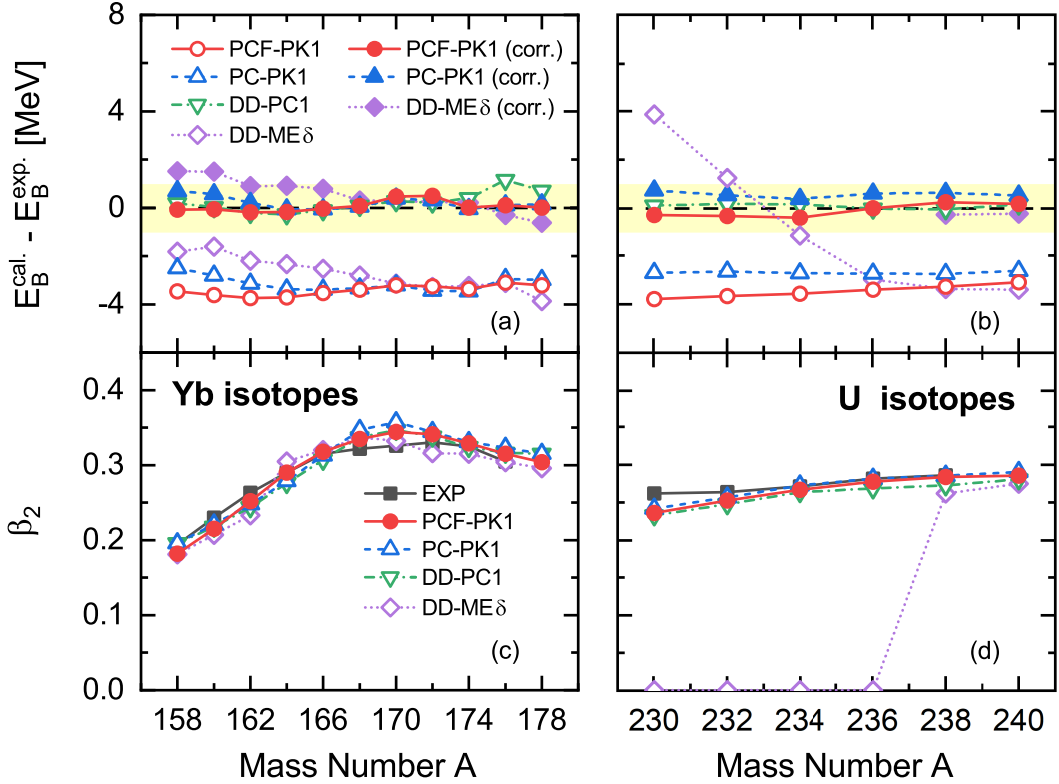}
  \caption{(Color online).
  Panels (c) and (d): Deviations of the calculated binding energies from the experimental data \cite{Wang_CPC41_030003} for Yb and U  isotopes obtained by PCF-PK1, PC-PK1, DD-PC1, and DD-ME$\delta$. The solid (open) symbols denote the values with (without) the rotational correction energies.
  Panel (c) and (d): Ground-state deformations of Yb and U isotopes given by PCF-PK1, PC-PK1, DD-PC1, and DD-ME$\delta$ in comparison with the data \cite{Pritychenko_ADaNDT107_1}.
  }
  \label{fig:Yb_U}
\end{figure*}

The ground-state binding energies and quadrupole deformations for the well-deformed Yb and U isotopes calculated with the density functional PCF-PK1, and the results are compared with the ones by PC-PK1, DD-PC1, and DD-ME$\delta$ as well as the data.
In Figs. \ref{fig:Yb_U}(a) and \ref{fig:Yb_U}(b), the deviations of the calculated binding energies from the experimental data \cite{Wang_CPC41_030003} are depicted.
Without the rotational correction energies, the PCF-PK1 systematically underestimates the binding energies about 3---4 MeV for both Yb and U isotopes; similar to PC-PK1.
For DD-ME$\delta$, it also underestimates the binding energies in most cases, but the deviations exhibit a clear isospin dependence behavior.
In particular for the lighter U isotopes, the DD-ME$\delta$ even overestimates the binding energies.
The DD-PC1 can well describe the binding energies with the deviations less than 1 MeV, because almost all nuclei calculated here were used in the fitting procedure of DD-PC1.

One should keep in mind that for deformed nuclei, due to the rotational symmetry breaking, the correction energies associated with the restoration of the rotational symmetry should be considered in addition to the mean-field energies.
Therefore, similar to Ref.~\cite{Zhao2010Phys.Rev.C54319}, here the rotational correction energies are calculated with the cranking approximation \cite{Girod_NPA330_40}.
After taking into account the rotational correction energies, the calculated results by PCF-PK1 and PC-PK1 are in good agreement with the experimental  binding energies for both Yb and U isotopes, and the discrepancies are within 1 MeV.
For DD-ME$\delta$, the deviations from the data for the Yb isotopes are slightly larger because of the apparent isospin dependence.
Moreover, the rotational correction energies have not been calculated for the nuclei from $^{230}$U to $^{236}$U with DD-ME$\delta$, because they are predicted to be spherical nuclei in the calculations.

In Figs. \ref{fig:Yb_U}(c) and \ref{fig:Yb_U}(d), it depicts the quadrupole deformations of the ground states for the Yb and U isotopes obtained by PCF-PK1, PC-PK1, DD-PC1, and DD-ME$\delta$ in comparison with data \cite{Pritychenko_ADaNDT107_1}.
Generally speaking, the deformations are well reproduced by all density functionals except for the DD-ME$\delta$ results of U isotopes from $^{230}$U to $^{236}$U.
As seen in Fig \ref{fig:d2p}, the DD-ME$\delta$ predicts a large spurious shell closure at $Z=92$, and this results in a spherical shape for the ground states of $^{230-236}$U.

\section{Summary}\label{sec:summary}

In summary, a new density-dependent point-coupling covariant density functional PCF-PK1 has been developed, in which the exchange terms of the four-fermion terms are taken into account with the Fierz transformation.
The new density functional PCF-PK1 contains 14 independent parameters in the mean-field channel, where 10 parameters are determined by the empirical saturation properties of nuclear matter and pseudo-data obtained from the \emph{ab initio} calculations.
The remain 4 parameters are optimized by fitting to the selected observables of 60 spherical nuclei including the binding energies, charge radii, and two-nucleon separation energies.
The performance of PCF-PK1 is illustrated with properties of the infinite nuclear matter and finite nuclei including the ground-state properties and the Gamow-Teller resonances.

For nuclear matter, the most prominent feature of the PCF-PK1 results is the large Dirac mass ($0.80 M$) and Landau mass ($0.85 M$), which is associated with the high level densities around the Fermi surface in finite nuclei.
It should be noted that the large Dirac mass ($0.80 M$) here does not worsen the description of the spin-orbit splittings in finite nuclei due to the inclusion of the tensor couplings in the functional, which also contribute to the spin-orbit potential.

For the spherical nuclei, the PCF-PK1 results can reproduce the experimental binding energies and charge radii quite well.
The results of two-proton shell gaps of the $N=82$ and $126$ isotones illustrate that the PCF-PK1 eliminates the spurious shell closures at $Z=58$ and $92$, which commonly exist in many relativistic density functionals.

Apart from the ground-states properties, the Gamow-Teller resonances of $^{48}$Ca, $^{90}$Zr, and $^{208}$Pb have also been calculated with the relativistic random-phase approximation.
Without any adjustable parameters, the PCF-PK1 reproduces the experimental excitation energies quite well.
This clearly demonstrates that a self-consistent description for the Gamow-Teller resonances can be achieved with present localized exchange terms.

For the deformed nuclei, the reliability of PCF-PK1 is illustrated by taking Yb and U isotopes as examples.
The quadrupole deformations are well reproduced by the PCF-PK1.
After taking into account the rotational correction energies, the PCF-PK1 results reproduce the experimental binding energies for the Yb and U isotopes within 1 MeV.

\section*{Acknowledgments}
The authors thank P. Ring for helpful discussions and suggestions and H.Z. Liang for providing the relativistic RPA code.
This work is supported by the National Key R\&D Program of China (Contracts No. 2018YFA0404400 and 2017YFE0116700), the National Natural Science Foundation of China (Grants No. 12070131001, 11875075, 11935003, 11975031, and 12141501), the China Postdoctoral Science Foundation under Grant No. 2020M670013, the IBS grant funded by the Korean government No. IBS-R031-D1 (Q.Z.), and the High-performance Computing Platform of Peking University.


%

\end{document}